\documentclass[runningheads]{llncs}

\usepackage[utf8]{inputenc}
\usepackage[T1]{fontenc}
\usepackage{lmodern}
\usepackage{textcomp}

\usepackage{amsmath}
\usepackage{amsfonts}
\usepackage[linesnumbered,ruled,vlined]{algorithm2e}
\usepackage{siunitx} 

\usepackage{graphicx} 
\graphicspath{{./images/}} 
\usepackage{caption}       
\usepackage{subcaption}    
\usepackage{booktabs}
\usepackage{multirow}
\usepackage{tikz}          
\usetikzlibrary{arrows}    

\usepackage{listings}
\usepackage{parcolumns}

\usepackage{xspace}
\usepackage{paralist}

\usepackage{cite}

\usepackage[hidelinks]{hyperref} 

\usepackage[all]{hypcap}
\usepackage[nameinlink,capitalise]{cleveref} 

\usepackage{todonotes}
\presetkeys{todonotes}{inline}{}

\newcommand{\eg}{\textit{e.g.,}}

\newcommand{\ie}{\textit{i.e.,}}

\newcommand{\etal}{\textit{et al.}}

\newcommand{\wrt}{\textit{w.r.t.}}

\newcommand{\evosuite}{\textsc{EvoSuite}}
\newcommand{\newcrossover}{\textit{HMX}}
\newcommand{\sbx}{\textit{SBX}}
\newcommand{\dynamosa}{\textit{DynaMOSA}}
\newcommand{\crossoverkey}{\texttt{multi\_lev\-el\_crossover}}

\newcommand{\pvalue}{p\text{-value}}
\newcommand{\atwelve}{\mathrm{\hat{A}}_{12}}

\newcommand{\nrun}{\num{100}}
\newcommand{\toruns}{\num{23200}}
\newcommand{\ncomps}{\num{9}}

\tikzstyle{mybox} = [draw=black, thick, rectangle, inner ysep=5pt, inner xsep=5pt] 

\begin{document}

\title{Hybrid Multi-level Crossover for Unit Test Case Generation}
\titlerunning{Hybrid Multi-level Crossover for Unit Test Case Generation}

\author{
  Mitchell Olsthoorn\orcidID{0000-0003-0551-6690} \and 
  Pouria Derakhshanfar\orcidID{0000-0003-3549-9019} \and 
  Annibale Panichella\orcidID{0000-0002-7395-3588}
}
\authorrunning{
  M. Olsthoorn \and
  P. Derakhshanfar \and
  A. Panichella
}

\institute{Delft University of Technology, Delft, The Netherlands\\
  \email{M.J.G.Olsthoorn@tudelft.nl}, 
  \email{P.Derakhshanfar@tudelft.nl}, 
  \email{A.Panichella@tudelft.nl}
}

\maketitle

\begin{abstract}
State-of-the-art search-based approaches for test case generation work at test case level, where tests are represented as sequences of statements.
These approaches make use of genetic operators (\ie{} mutation and crossover) that create test variants by adding, altering, and removing statements from existing tests.
While this encoding schema has been shown to be very effective for many-objective test case generation, the standard crossover operator (single-point) only alters the structure of the test cases but not the input data.
In this paper, we argue that changing both the test case structure and the input data is necessary to increase the genetic variation and improve the search process. 
Hence, we propose a hybrid multi-level crossover (\newcrossover{}) operator that combines the traditional test-level crossover with data-level recombination.
The former evolves and alters the test case structures, while the latter evolves the input data using numeric and string-based recombinational operators.
We evaluate our new crossover operator by performing an empirical study on more than 100 classes selected from open-source Java libraries for numerical operations and string manipulation.
We compare \newcrossover{} with the single-point crossover that is used in \evosuite{} \wrt{} structural coverage and fault detection capability.
Our results show that \newcrossover{} achieves a statistically significant increase in \SI{30}{\percent} of the classes up to \SI{19}{\percent} in structural coverage compared to the single-point crossover. Moreover, the fault detection capability improved up to \SI{12}{\percent} measured using strong mutation score.

\end{abstract}

\section{Introduction}
\label{sec:introduction}

Genetic operators are a fundamental component of evolutionary search-based test case generation algorithms.
These operators create variation in the test cases to help the search process explore new possible paths.
The main genetic operators are mutation, which makes changes to a single test case, and crossover, which exchanges information between two test cases.

Over the years, related work has used three types of encoding schemata to represent test cases for search algorithms, namely data-level, test case-level, and test suite-level.
These schemata typically implement genetic operators at the same level as the encoding.
For example, the crossover operator at the data-level exchanges data between two input vectors~\cite{McMinn2004}.
The test case-level crossover exchanges statements between two parent test cases~\cite{tonella2004evolutionary}.
Lastly, the test suite-level crossover swaps test cases within two test suites~\cite{Fraser2011}.
Recent studies have shown that the test case-level schema combined with many-objective (MO) search is the most effective at generating test cases with high coverage~\cite{panichella2017automated, Campos2018}.
 
The current many-objective approaches use the single-point crossover to recombine groups of statements within test cases.
Test cases consist of both test structures (method sequences) and test data~\cite{tonella2004evolutionary}.
Hence, the crossover operator only changes the test structure and simply copies over the corresponding input data.
Therefore, input data has to be altered by the mutation operator, usually with a small probability.

In this paper, we argue that better genetic variation can be obtained by designing a crossover operator that alters the structure of the test cases and also the input data by creating new data that is in the neighborhood of the parents' data.
To validate this hypothesis, we propose a new operator, called Hybrid Multi-level Crossover (\newcrossover{}), that combines different crossover operators on multiple levels.
We implement \newcrossover{} within \evosuite{}~\cite{Fraser2011}, the state-of-the-art unit-test generation tool for Java.

To evaluate the effectiveness of our operator, we performed an empirical study where we compare \newcrossover{} with the single-point crossover used in \evosuite{}, a state-of-the-art test case generation tool for Java, \wrt{} structural coverage and fault detection capability.
To this aim, we build a benchmark with 116 classes from the Apache Commons and Lucene Stemmer projects, which include classes for numerical operations and string manipulation.

Our results show that \newcrossover{} achieves higher structural coverage for \textasciitilde\SI{30}{\percent} of the classes in the benchmark.
On average, \newcrossover{}, covered \SI{6.4}{\percent} and \SI{7.2}{\percent} more branches and lines than our baseline, respectively (with a max improvement of \SI{19.1}{\percent} and \SI{19.4}{\percent}).
Additionally, the proposed operator improved the fault detection capability in \textasciitilde\SI{25}{\percent} of the classes with an average improvement of \SI{3.9}{\percent} (max. \SI{14}{\percent}) and \SI{2.1}{\percent} (max. \SI{12.1}{\percent}) for weak and strong mutation, respectively.

In summary, we make the following contributions:
\begin{enumerate}
	\item A novel crossover that works at both test case and input data-level to increase genetic variation in the search process. The data-level recombination combines multiple different techniques depending on the data type.
	\item An open-source implementation of our operator in \evosuite{}.
	\item A full replication package containing the results and the analysis scripts~\cite{mitchell_olsthoorn_2021_5102597}.
\end{enumerate}

The outline for the remainder of this paper is as follows. 
\cref{sec:background} explains the fundamental concepts used in the paper. 
\cref{sec:approach} introduces our new crossover operator, called \newcrossover{}, and breaks down how it works. 
\cref{sec:study} sets out our research questions and describes the setup of our empirical study. 
\cref{sec:results} details our results and highlights our findings. 
\cref{sec:validity} discusses the threats to validity and \cref{sec:conclusion} draws conclusions and identifies possible directions for future work.

\section{Background and Related Work}
\label{sec:background}

\textbf{Search-based unit test generation.}
Prior studies introduced search-based software test generation (SBST) approaches utilizing meta-heuristics (\eg{} genetic algorithm) to automate test generation for different testing levels~\cite{McMinn2004}, such as unit~\cite{Fraser2011}, integration~\cite{derakhshanfar2020integ}, and system-level testing~\cite{arcuri2019restful}.
Search-based unit-test generation is one of the widely studied topics in this field, where a search process generates tests fulfilling various criteria (\eg{} structural coverage, mutation score) for a given class under test (CUT).
Studies have shown that these techniques are effective at achieving high code coverage~\cite{Campos2018, Panichella2018a} and fault detection~\cite{Almasi2017}. 

\medskip
\textbf{Single-objective unit test generation.}
Single-objective techniques specify one or more fitness functions to guide the search process towards covering the search targets according to the desired criteria.
Rojas \etal{} \cite{rojas2015whole} proposed an approach that aggregates all of the fitness functions for each criterion using a weighted sum scalarization and performs a single-objective optimization to generate tests.
Additionally, Gay~\cite{gay2017generating} empirically showed that combining different criteria in a single-objective leads to detect more faults compared to using each criterion separately.

\medskip
\textbf{Dynamic many-objective sorting algorithm (\dynamosa{}).}
In contrast with single-objective unit test generation, Panichella \etal{} have proposed a many-objective evolutionary-based approach, called \dynamosa{}~\cite{panichella2017automated}. This approach considers each coverage targets from multiple criteria as an independent search objective.
\dynamosa{} utilizes the hierarchy of dependencies between different coverage targets (\eg{} line, branch, mutants) to select the search objectives during the search dynamically.
Moreover, recent work~\cite{panichella2018incremental} introduced a multi-criteria variant of \dynamosa{} that extends the idea of dynamic selection of the targets, based on an enhanced hierarchical dependency analysis.
This recent study showed that this multi-criteria variant outperforms single-objective search-based unit test generation \wrt{} structural and mutation coverage and, therefore, can achieve a higher fault detection rate.
These results have also been confirmed independently by Campos \etal~\cite{Campos2018}.
Consequently, \dynamosa{} is currently used as the default algorithm in \evosuite{}.

\medskip
\textbf{Crossover operator.}
Like any other evolutionary-based algorithms, all variations of \dynamosa{} need crossover and mutation operators for evolving the individuals in the current population to generate the next population. 
Since \dynamosa{} encodes tests at a test case-level, the mutation operator alters statements in a selected test case according to a given \textit{mutation probability}.
This search algorithm uses the single-point crossover to recombine two selected individuals (parents) into new tests (offspring) for the next generation.
This crossover operator randomly selects two positions in the selected parents and split them into two parts.
Then, it remerges each part with the opposing part from the other parent. A more detailed explanation of this operator is available in \cref{sec:approach}.

While the single-point crossover brings diversity to the structure of the generated test cases, it does not work at the data-level (\ie crossover between the test inputs).
Hence, this study introduces a hybrid multi-level crossover, called \newcrossover{}, for the state-of-the-art in search-based unit test generation.
\section{Approach}
\label{sec:approach}

This section details our new crossover operator, called Hybrid Multi-level Crossover (\newcrossover{}).
This operator combines the traditional \textit{single-point} test case-level crossover with multiple data-level crossovers.

\begin{algorithm}[t]
\footnotesize
\DontPrintSemicolon
\KwIn{Two parent test cases $P_1$ and $P_2$\;}
\KwOut{Two offspring test cases $O_1$ and $O_2$\;}
\Begin{
$O_1, O_2 \gets$ SINGLE-POINT-CROSSOVER($P_1$, $P_2$)\;

\tcp{Constructor data store}
$C_1 \gets$ Map<signature, constructor[ ]> \tcp{For $P_1$}
$C_2 \gets$ Map<signature, constructor[ ]> \tcp{For $P_2$}

\tcp{Method data store}
$M_1 \gets$ Map<signature, method[ ]> \tcp{For $P_1$}
$M_2 \gets$ Map<signature, method[ ]> \tcp{For $P_2$}

\ForAll{($S_1$, $S_2$),  in $S_1 \in O_1$ and $S_2 \in O_2$}{
	\If{SIGNATURE($S_1$) $==$ SIGNATURE($S_2$)}{
		\If{$S_1$ is constructor}{
			$C_1$[SIGNATURE($S_1$)].add($S_1$)\;
			$C_2$[SIGNATURE($S_2$)].add($S_2$)\;
		}
		\ElseIf{$S_1$ is method}{
			$M_1$[SIGNATURE($S_1$)].add($S_1$)\;
			$M_2$[SIGNATURE($S_2$)].add($S_2$)\;
		}
	}
}

\ForEach{$SIG \in C_1.keys \cup C_2.keys$}{
	\tcp{choose random constructor with same signature}
	$S_1 \gets random.choice(C1[SIG])$\;
	$S_2 \gets random.choice(C2[SIG])$\;
	$O_1, O_2 \gets$ DATA-CROSSOVER($O_1$, $O_2$, PARAMS($S_1$), PARAMS($S_2$))\;
}
\ForEach{$SIG \in M_1.keys \cup M_2.keys$}{
	\tcp{choose random method with same signature}
	$S_1 \gets random.choice(M1[SIG])$\;
	$S_2 \gets random.choice(M2[SIG])$\;
	$O_1, O_2 \gets$ DATA-CROSSOVER($O_1$, $O_2$, PARAMS($S_1$), PARAMS($S_2$))\;
}
\Return{$O_1$, $O_2$}\;
}
\caption{\newcrossover{}: hybrid multi-level crossover\label{algo:hmx}}
\end{algorithm}

\cref{algo:hmx} outlines the pseudo-code of our crossover operator.
\newcrossover{} first performs the traditional \textit{single-point} crossover at line 2.
The \textit{single-point} crossover is chosen for the test case-level operator as previous studies have shown that it is effective in producing a variation in the population over time~\cite{tonella2004evolutionary}.
It is also the default crossover operator used in the state-of-the-art test case generation tool \evosuite{}~\cite{tonella2004evolutionary}.
This operator takes two parent test cases as input and selects a random point among the statements within the parents test cases.
The parents are then split at this point, and their resulting parts are then recombined with its opposing part of the other parent to produce two new offspring test cases. 
Since these offspring test cases use a random crossover point, they might contain incomplete sequences of statements (\eg{} missing variable definition) and, therefore, will not compile.
To make the crossover more effective, these broken references are fixed by introducing new random variable definitions that match the type of the broken reference~\cite{Fraser2011}.
Lines 3-22 contain the selection logic of the data-level crossover.
Unlike the test case-level crossover, the data-level crossover can not be applied to every combination of input data.
Performing the crossover on input data with different types (\eg{} strings and numbers) would not produce any meaningful output as there is no logical way to combine these dissimilar types.
Furthermore, we should not perform a crossover on two identical data types from different methods.
If the data-level crossover would be applied to parameters of the same type that belong to different methods, it could produce offspring that are farther from the desired objective than the original.
Hence, the algorithm has to select which combinations of input data are compatible.
\newcrossover{} achieves this by selecting compatible functions (\ie{} constructors and methods calls) and applying the crossover pairwise to the function's parameters.

In lines 3-6, two pairs of maps are created that store the compatible functions for each parent for both constructors and methods.
Each map stores a list of functions that share the same signature; The signature is the key of the map, and the functions are the values.
The signature of the function is a string derived from the class name, function name, parameters types, and return type using the following format:
\begin{equation*}
	{\scriptstyle\texttt{CLASS\_NAME|FUNCTION\_NAME(PARAM1\_TYPE, PARAM2\_TYPE, \ldots)RETURN\_TYPE}}
\end{equation*}

In lines 7-14, \newcrossover{} loops over all combinations of statements $S_1$ and $S_2$ in the offspring produced by the single-point crossover.
For each combination, it checks if the signatures of the two functions match (line 8).
If both statements are either constructors or methods, they are stored in their corresponding map with the signature as a key in lines 10-11 and 13-14, respectively.
Note that if the test case contains constructor or method calls for other classes than the CUT, these are also considered by the selection of compatible functions.
For example, additional objects (\eg{} strings, lists) might be needed as an input argument to one of the CUT's functions.

When all possible matching functions have been found, the operator loops through the signatures of the two function types separately in lines 15-18 and 19-22.
For each signature, \newcrossover{} selects a random function instance matching the signature from each parent.
The operator then performs the data-level crossover on the parameters of these two randomly selected functions in lines 18 and 22.
For each signature in the map, \newcrossover{} only selects one function instance per parent to proceed with the genetic recombination.

The data-level recombination pairwise traverses the parameters of the two compatible functions selected in lines 16-17 (for constructors) and 20-21 (for methods).
For each pair of parameters, \cref{algo:hmx} checks their types and determines if they are numbers or strings, the two supported types of \newcrossover{}.
If the two parameters are numbers (\ie{} byte, short, int, long, float, double, boolean, and char), the operator applies the \textit{Simulated Binary Crossover} (SBX), which is described in \cref{sec:sbx}.
If the parameters are strings, it applies the string crossover described in \cref{sec:string-crossover}.
Lastly, in line 23, \newcrossover{} returns the produced offspring.

\begin{lstlisting}[
caption=Parent 1,
label=lst:parent1,
frame=tb,
numbers=left,
language=Java,
basicstyle=\ttfamily\footnotesize,
xleftmargin=2em,
framexleftmargin=2em
]{Example1}
@Test
public void test1() {
  Fraction f0 = new Fraction(2, 3);
  Fraction f1 = new Fraction(2, -1);
  f0.divideBy(f1);
  f0.add(Fraction.ZERO);
}
\end{lstlisting}

\begin{lstlisting}[
caption=Parent 2,
label=lst:parent2,
frame=tb,
numbers=left,
language=Java,
basicstyle=\ttfamily\footnotesize,
xleftmargin=2em,
framexleftmargin=2em
]{Name}
@Test
public void test2() {
  Fraction f0 = new Fraction(3, 1);
  Fraction f1 = new Fraction(1, 3);
  f0.add(f1);
  f0.pow(2.0);
}
\end{lstlisting}

To provide a practical example, let us consider the two parent test cases in \cref{lst:parent1,lst:parent2}.
Both parent 1 and parent 2 contain two invocations of the \texttt{Fraction} constructor. Since these constructors share the same signature: ${\footnotesize\texttt{Fraction|<init>(int, int)Fraction}}$; they are compatible.
Similarly, the method \texttt{add} of the \texttt{Fraction} class is present in both parents, with the same signature: ${\footnotesize\texttt{Fraction|add(Fraction)V}}$; and are compatible, as well.
In contrast, for example, method \texttt{divideBy}, in parent 1, and method \texttt{add}, in parent 2, are not compatible since their signatures are different.

\subsection{Simulated Binary Crossover}
\label{sec:sbx}

The \textit{Simulated Binary Crossover} (SBX) is a recombination operator commonly used in numerical  problems with numerical decision variables and fixed-length chromosomes.
It has been shown that Evolutionary Algorithms (EAs) that use this crossover operator produce better results compared to traditional numerical crossover operators~\cite{deb2007self}.
The equation below outlines the algorithm of \sbx{}:
\begin{align}
	\label{eqn:u}
	u &= rand_{u}\\
	\label{eqn:beta}
	\beta &= 
  	\begin{cases} 
   		{2 \cdot u}^{1 / (\eta_c + 1)} & \text{if } u < 0.5\\
   		1 & \text{if } u = 0.5\\
   		{\frac{0.5}{1.0 - u}}^{1 / (\eta_c + 1)} & \text{if } u > 0.5\\
  	\end{cases}\\
  	\label{eqn:b}
  	b &= rand_b\\
  	\label{eqn:v}
	v &= 
  	\begin{cases} 
   		((v_1 - v_2) \cdot 0.5) - (\beta \cdot 0.5 \cdot \lvert v_1 - v_2 \rvert) & \text{if } b = true \\
   		((v_1 - v_2) \cdot 0.5) + (\beta \cdot 0.5 \cdot \lvert v_1 - v_2 \rvert) & \text{if } b = false \\
  	\end{cases}
\end{align}
where $v$ (\cref{eqn:v}) is the new value of parameter $v_1$, $v_1$ is the original value of the parameter, and $v_2$ is the value of the opposing parameter (the corresponding parameter from the matched function).
$\eta_c$ is the \textit{distribution index} and it measures how close the new values should be to original values (proximity).
For \newcrossover{}, this variable is set to 2.5 as this is within the recommended range [2;5]~\cite{deb2007self}.
\sbx{} first creates a random \textit{uniform} variable $u$ (\cref{eqn:u}), which is used to select one of three strategies for $\beta$.
This scaling variable $\beta$ (\cref{eqn:beta}), is used to scale an offset.
This offset is either subtracted or added depending on the random \textit{boolean} variable $b$.
In general, \sbx{} generates new values centered around the original parents, either in between the parents' values (contracting) or outside this range (expending) depending on the value of $u$.
The algorithm is performed on both matching parameters, and the resulting new values are used as a replacement of the original values.

As an example, consider the two compatible constructors \texttt{Fraction(2,3)} (line 3 in \cref{lst:parent1}) and \texttt{Fraction(1,3)} (line 4 in \cref{lst:parent2}).
The \sbx{} recombination operator is applied for the following pairwise combinations: (2, 1) and (3, 3).
To calculate the new value of the first element of the first pair, $v_1 = 2$ and $v_2 = 1$.
Similarly, the second element can be calculated by switching the values of $v_1$ and $v_2$.
The same procedure can be applied to calculate the new values of the second pair.

\subsection{String Crossover}
\label{sec:string-crossover}

The single-point string crossover is used to exchange information between two string parameters of matching functions~\cite{McMinn2004}.
By recombining parts of each string, it makes it possible for promising substrings to collect together.
The operator achieves this by picking two random numbers, $0 \leq x_i <$ length($x$) and $0 \leq y_i <$ length($y$) for both strings, respectively.
It then recombines the two strings by concatenating the substrings in the following way: $x = x[:x_i] \;||\; y[y_i:]$ and $y = y[:y_i] \;||\; x[x_i:]$.

For example, given the following string $x = "lorem"$ and $y = "ipsum"$ and the random variables $x_i = 1$ and $y_i = 3$, the new values will be: $x = "lom"$ and $y = "ipsurem"$. 

\section{Empirical Study}
\label{sec:study}

To assess the impact of \newcrossover{} on search-based unit test generation, we perform an empirical evaluation to answer the following research questions:

\begin{itemize}
    \item[\textbf{RQ1}] \textit{To what extent does \newcrossover{} improve structural coverage compared to the single-point crossover?}
	\item[\textbf{RQ2}] \textit{How does \newcrossover{} impact the fault-detection capability of the generated tests?}
\end{itemize}

\medskip
\textbf{Benchmark.}
\begin{table*} [t]
	\centering
	\caption{Projects in our empirical study. \# indicates the number of CUTs. \texttt{cc} indicates the cyclomatic complexity of CUTs. \texttt{$\sigma$} indicates the standard deviation. \texttt{min} and \texttt{max} indicate the minimum and maximum value of the metric, respectively. Also, $\overline{\text{str-par}}$ and $\overline{\text{nr-par}}$ are the average number of string and number input parameters for the selected CUTs.}
	\label{tab:projects}
	\begin{tabular}{ l r | r@{\hskip 0.08in}r@{\hskip 0.08in}r@{\hskip 0.08in}r | r@{\hskip 0.08in}r@{\hskip 0.08in}r@{\hskip 0.08in}r | r@{\hskip 0.08in}r@{\hskip 0.08in}r@{\hskip 0.08in}r }
\hline 
\textbf{Project} & \textbf{\#} & \multicolumn{4}{c}{\textbf{CCN}} & \multicolumn{4}{c}{\textbf{String parameter}} & \multicolumn{4}{c}{\textbf{Number parameter}} \\ 
  &   & $\overline{\text{cc}}$ & $\sigma$ & min & max & $\overline{\text{str-par}}$ & $\sigma$ & min & max & $\overline{\text{nr-par}}$ & $\sigma$ & min & max \\ 
\hline 
CLI & 4 &1.7 &0.9 &3.0 & 1.1 &14.5 &14.2 &34.0 & 4.0 &8.5 &13.7 &29.0 & 1.0 \\ 
Geometry & 13 &1.8 &0.4 &2.5 & 1.2 &3.4 &5.5 &21.0 & 1.0 &10.2 &6.7 &21.0 & 1.0 \\ 
Lang & 34 &3.0 &1.6 &7.4 & 1.1 &17.4 &36.7 &209.0 & 1.0 &26.6 &48.3 &249.0 & 1.0 \\ 
Logging & 1 &3.0 &- &3.0 & 3.0 &6.0 &- &6.0 & 6.0 &3.0 &- &3.0 & 3.0 \\ 
Math & 27 &2.9 &1.6 &7.7 & 1.1 &2.5 &1.8 &9.0 & 1.0 &10.0 &10.5 &45.0 & 1.0 \\ 
Numbers & 5 &2.8 &1.1 &4.5 & 1.6 &1.4 &0.9 &3.0 & 1.0 &31.6 &33.5 &89.0 & 4.0 \\ 
RNG & 4 &3.3 &1.4 &5.0 & 1.7 &2.2 &2.5 &6.0 & 1.0 &2.0 &1.4 &4.0 & 1.0 \\ 
Stemmer & 16 &1.0 &0.0 &1.0 & 1.0 &0.0 &0.0 &0.0 & 0.0 &0.0 &0.0 &0.0 & 0.0 \\ 
\hline 
\end{tabular}

\end{table*}
For this study, we selected the CUTs from the \textsc{Apache Commons} and \textsc{Snowball Stemmer} libraries. The former is a commonly-used project containing reusable Java components for several applications \footnote{\url{https://commons.apache.org}}. The latter is a well-known library for stemming strings, which is part of the \textsc{Apache Lucene} \footnote{\url{https://github.com/weavejester/snowball-stemmer}}.
As described in \cref{sec:approach}, \newcrossover{} brings more advantages for search-based test generation in projects that utilize strings and numbers.
Hence, to show the effect of this new crossover operator, we selected \num{100}~classes from \ncomps{} components in \textsc{Apache Commons} that have numeric and string input data:
\begin{inparaenum}[(i)]
    \item \textsc{Math} a library of lightweight, self-contained mathematics and statistics components;
    \item \textsc{Numbers} includes utilities for working with complex numbers;
    \item \textsc{Geometry} provides utilities for geometric processing;
    \item \textsc{RNG} a library of Java implementations of pseudo-random generators;
    \item \textsc{Statistics} a project containing tools for statistics;
    \item \textsc{CLI} an API processing and validating a command line interface;
    \item \textsc{Text} a library focused on algorithms working on strings; 
    \item \textsc{Lang} contains extra functionality for classes in \texttt{java.lang}; and 
    \item \textsc{Logging} an adapter allowing configurable bridging to other logging systems.
\end{inparaenum}

In addition, we added the main 16 classes in \textsc{Snowball Stemmer} to the benchmark, as these focus on string manipulation and were previously used in former search-based unit test generation studies~\cite{panichella2015reformulating}.

Due to the large number of classes in the selected \textsc{Apache Commons} components, we used \textsc{CK}~\cite{ck}, a tool that calculates the method-level and class-level code metrics in Java projects using static analysis.
We collect the Cyclomatic Complexity (CC) and type of input parameters for each method in the selected \ncomps{} components.
Using the collected information, we filter out the classes that do not have methods accepting strings or numbers (integer, double, long, or float) as input parameters. 
Then, we sort the remaining classes according to their average CC and pick the top \num{100}~cases for our benchmark.
Table \ref{tab:projects} reports CC, number of string, and number arguments for each project used in this study.
By doing a preliminary run of \evosuite{} on the \num{116}~selected classes, we noticed that this tool fails to start the search process in 9 of the CUTs.
These failures stem from an issue in the underlying test generation tool \evosuite{}.
The tool fails to gather a critical statistic (\ie{} TOTAL\_GOALS) for these runs in both the baseline and \newcrossover{}.
We also encountered 4 classes that did not produce any coverage for both the baseline and our approach.
Consequently, we filtered out these classes from the experiment and performed the final evaluation on 103 remaining classes.

\medskip
\textbf{Implementation.}
We implemented \newcrossover{} in \evosuite{}~\cite{Fraser2011}, which is the state-of-the-art tool for search-based unit test generation in Java.
By default, this tool uses the single-point crossover for test generation.
We have defined a new parameter \crossoverkey{} to enable \newcrossover{}.
Our Implementation is openly available as an artifact~\cite{mitchell_olsthoorn_2021_5102597}.

\medskip
\textbf{Preliminary Study.}
We performed a preliminary study to see how the probability of applying our data-level crossover influences the result.
The single-point test case-level crossover is applied with a predefined probability.
We experimented with how often the data-level crossover should be applied whenever the test case-level crossover was applied.
From the probabilities we tried (\ie{} 0.25, 0.50, 0.75, 1.00), we found out that always applying the data-level crossover when the test case-level crossover produced the best results according to statistical analysis.

\medskip
\textbf{Parameter Settings.}
We run each search process with \evosuite{}'s default parameter values.
As confirmed by prior studies~\cite{arcuri2013parameter}, despite the impact of parameter tuning on the search performance, the default parameters provide acceptable results. 
Hence, we run each search process with a two-minute search budget and set the population size to 50 individuals. 
Moreover, we use mutation with a probability of $1/n$ ($n =$ length of the generated test).
For both crossover operators that we used in this study (single-point crossover for the baseline and our novel \newcrossover{}), the crossover probability is $0.75$.
For the Simulated Binary Crossover (SBX), we used the \textit{distribution index} 
$\eta_c = 2.5$~\cite{deb2007self}.
The search algorithm is the multi-criteria DynaMOSA~\cite{panichella2018incremental}, which is the default one in \evosuite{} \texttt{v1.1.0}.

\medskip
\textbf{Experimental Protocol.}
We apply both default \evosuite{} with single-point crossover and \evosuite{} + \newcrossover{} to each of the selected CUTs in the benchmark.
To address the random nature of search-based test generation tools, we repeat each execution \nrun{}~times, with a different random seed, for a total number of \toruns{} independent executions.
We run our evaluation on a system with an AMD EPYC\texttrademark{} 7H12 using 240 cores running at \SI{2.6}{\giga\hertz}.
With each execution taking 5 minutes on average (\ie{} search, minimalization, and assertion generation), the total running time is \num{80.6}~days of sequential execution.

For our analysis, we report the average (median) results across the \nrun{} repeated runs.
To determine if the results (\ie{} structural code coverage and fault detection capability) of the two crossover operator are statistically significant, we use the unpaired Wilcoxon rank-sum test~\cite{conover1998practical} with a threshold of \num{0.05}.
The Wilcoxon test is a non-parametric statistical test that determines if two data distributions are significantly different.
Additionally, we use the Vargha-Delaney statistic~\cite{vargha2000critique} to measure the magnitude of the result, which determines how large the difference between the two operators is.

\section{Results}
\label{sec:results}

This section discusses the results of our study with the aim of answering the research questions formulated in \cref{sec:study}.
All differences in results in this section are presented in absolute differences (percentage points).

\subsection{Result for RQ1: Structural Coverage}
\label{sec:rq1}

\begin{figure*}[t]
\centering
\begin{subfigure}[b]{0.42\textwidth}
	\includegraphics[width=\columnwidth]{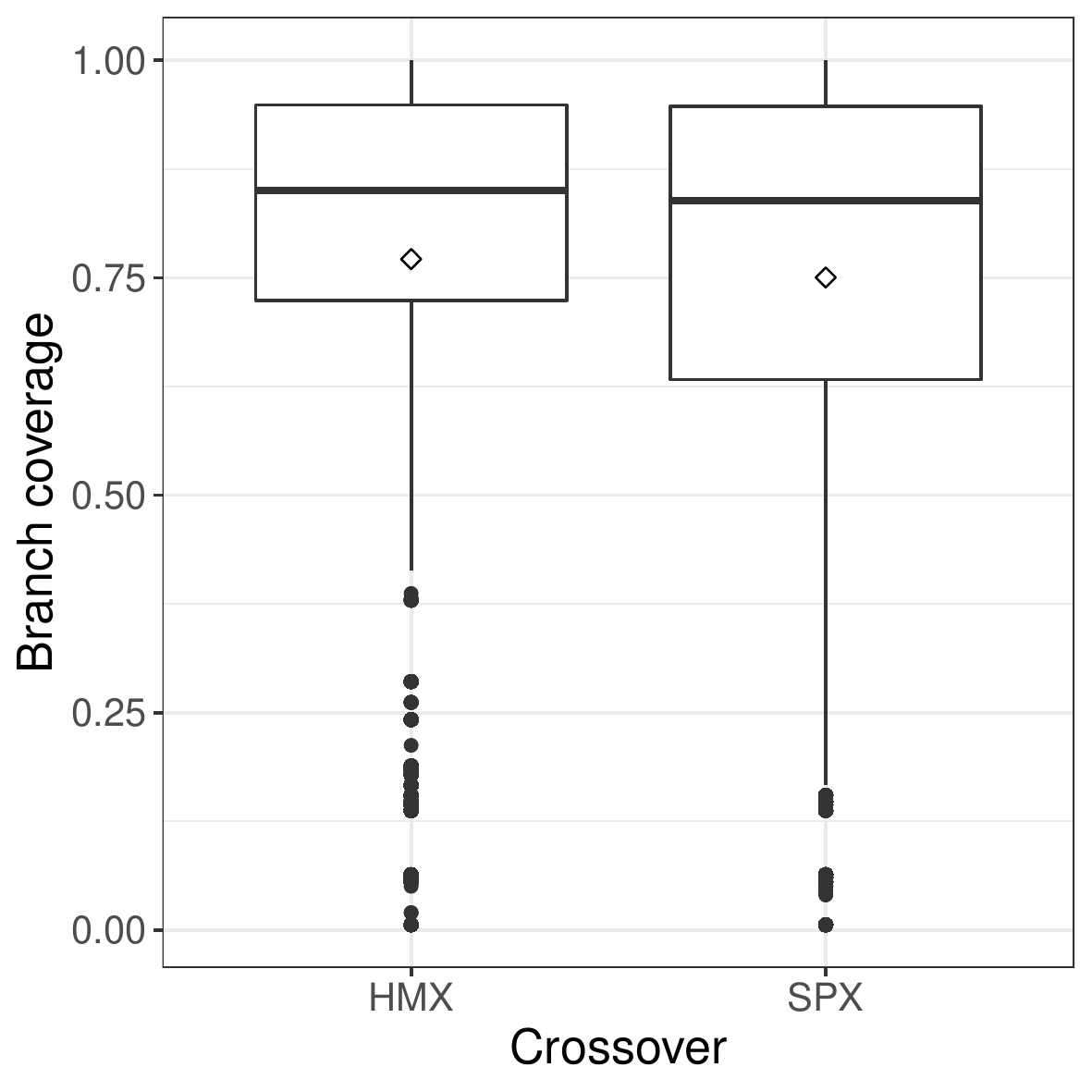}
	\caption{Branch Coverage}
	\label{fig:structural-branch}
\end{subfigure}
\begin{subfigure}[b]{0.42\textwidth}
	\includegraphics[width=\columnwidth]{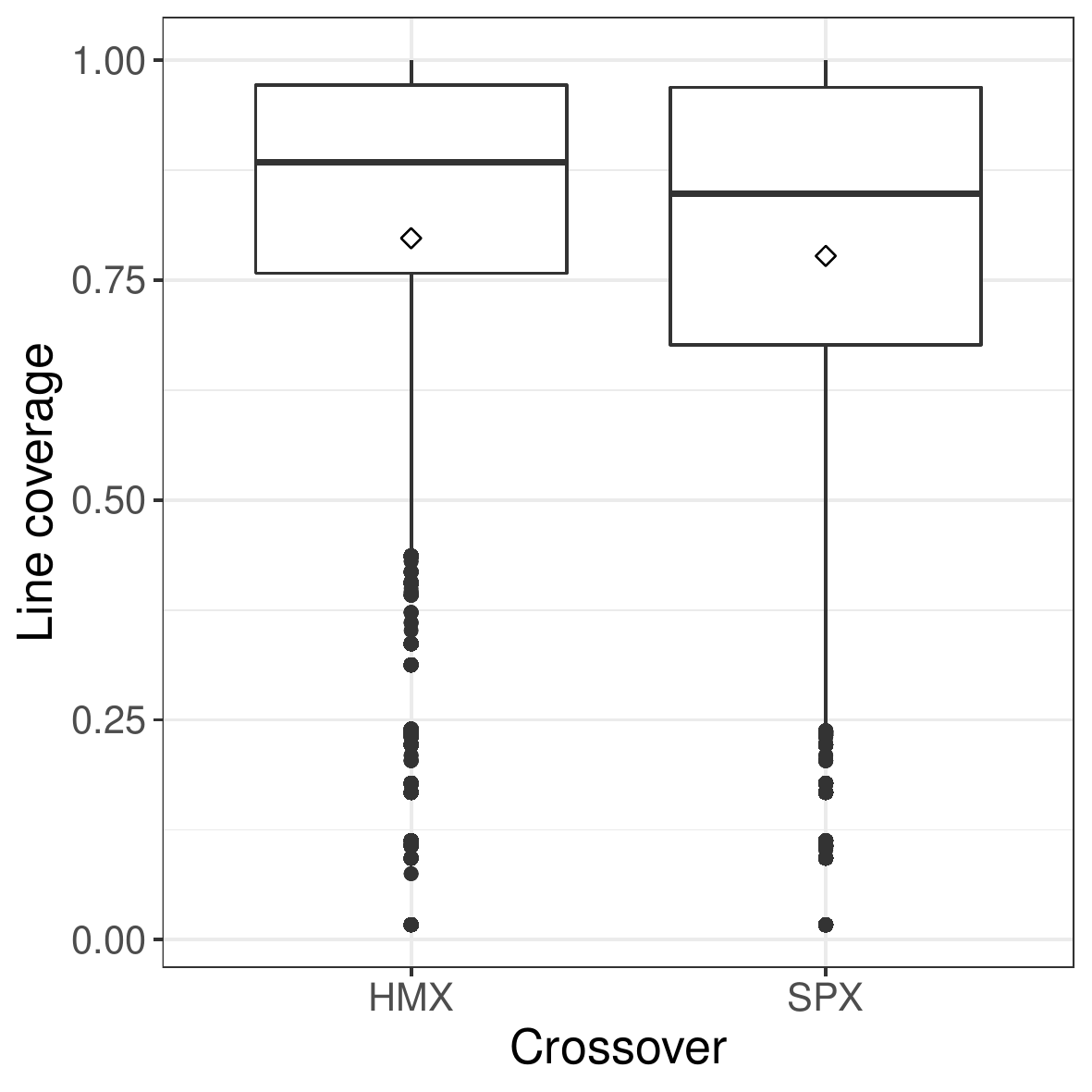}
	\caption{Line Coverage}
	\label{fig:structural-line}
\end{subfigure}
\caption{Boxplot of structural coverage comparing \newcrossover{} to the baseline SPX. The diamond point indicates the mean coverage of the benchmark.}
\label{fig:structural}
\end{figure*}

\cref{fig:structural} shows the structural coverage achieved by our approach, \newcrossover{}, compared to the baseline, SPX, on the benchmark.
In particular, \cref{fig:structural-branch} shows branch coverage and \cref{fig:structural-line} shows line coverage.
The boxplots show the median, quartiles, variability in the results, and the outliers for all classes together.
The diamond point indicates the mean of the results.

\cref{fig:structural-branch} and \cref{fig:structural-line} show that, on average, \newcrossover{} has higher $1^{st}$ quartile, median, mean, and $3^{rd}$ quartile values than the baseline, SPX, for both test metrics. 
On average, \newcrossover{} improves the branch coverage by +\SI{2.0}{\percent} and the line coverage by +\SI{1.9}{\percent}. The largest differences are visible for the lower whisker and for the first quartile (25th percentile). In particular, the differences for the lower whisker are around +20\% branch and line coverage when using \newcrossover{}; the improvements in the first quartile are around +10\% and +8\% for branch and line coverage, respectively. These results indicate that \newcrossover{} improves both line and branch coverage for some of the CUTs in our benchmark.
Finally, as we can see in both of the plots in \cref{fig:structural}, the variation in the results for \newcrossover{}, measured by the Interquartile Range (IQR), is smaller than for SPX. This observation shows that \newcrossover{} helps \evosuite{} to generate tests with a more stable structural coverage.

\begin{table}[t]
\centering
\caption{Statistical results of \newcrossover{} vs. SPX on structural coverage. \#Win indicates the number of times that \newcrossover{} is statistically better than SPX. \#Lose indicates the opposite. \#No diff. indicates that there is no statistical difference. Negl., Small, Medium, and Large denote the $\atwelve{}$ effect size.}
\scriptsize
\begin{tabular}{lc cccc c cccc  cc}
	\toprule
	\multirow{2}{*}{Metric} && \multicolumn{4}{c}{\#Win} && \multicolumn{4}{c}{\#Lose} && \multirow{ 2}{*}{\#No diff.} \\
	\cmidrule{2-6} \cmidrule{8-11}&& Negl. & Small & Medium & Large && Negl. & Small & Medium & Large \\
	\midrule
	Branch & & 2 & 5 & 3 & 22 && 0 & 1 & 0 & 0 && 70 \\
	Line & & 3 & 1 & 3 & 19 && 0 & 1 & 0 & 0 && 76 \\
	\bottomrule
\end{tabular}
\label{tab:structural-statistics}
\end{table}

\cref{tab:structural-statistics} shows the results of the statistical comparison between \newcrossover{} and the baseline, SPX, based on a $\pvalue{} \leq 0.05$.
\textit{\#Win} indicates the number of times that \newcrossover{} has a statistically significant improvement over SPX.
\textit{\#Equal} indicates the number of times that there is no statistical difference in the results between the two operators; \textit{\#Lose} indicates the number of times that \newcrossover{} has statistically worse results than SPX.
The \textit{\#Win} and \textit{\#Lose} columns also include the magnitude of the difference through the $\atwelve{}$ effect size, classified in \textit{Small}, \textit{Medium}, \textit{Large}, and \textit{Negligible}.

From \cref{tab:structural-statistics}, we can see that \newcrossover{} has a statistically significant non-negligible improvement in 30 and 23 classes for branch and line coverage, respectively.
For the branch coverage metric, \newcrossover{} improves with a large magnitude for 22 classes, medium for 3 classes, and small for 5 classes.
For line coverage, \newcrossover{} improves with a large magnitude for 19 classes, medium for 3 classes, and small for 1 class.
\newcrossover{} only loses in one case in comparison to the baseline for both branch and line coverage:  \texttt{StrSubstitutor} from the \texttt{Lang} project. However, in this case, the effect size is small (magnitude).

For branch coverage, we observe a maximum increase in coverage of +\SI{19.1}{\percent} for the \textit{finnishStemmer} class from the \texttt{Stemmer} project. For line coverage, the class with the maximum increase in coverage is \texttt{hungarianStemmer} (also from \texttt{Stemmer}) with an average improvement of +\SI{19.4}{\percent}.
Compared to the baseline, all classes in the \textsc{Snowball Stemmer} string manipulation library improve based on branch and line coverage with an average improvement of +\SI{11.4}{\percent} and +\SI{11.0}{\percent}, respectively.
For the \textsc{Apache Commons} library, \newcrossover{} significantly improves the branch and line coverage in 16 (9 string-related and 7 number-related) and 10 (6 string-related and 4 number-related) classes, respectively.

\vspace*{2mm}
\hspace*{-5mm}
\begin{tikzpicture}
\node [mybox] (box){%
\centering
\begin{minipage}{0.94\textwidth}
In summary, the proposed \newcrossover{} crossover operator achieves significantly higher (\textasciitilde\SI{30}{\percent} of the cases) or equal structural code coverage for unit test case generation compared to the baseline SPX.
\end{minipage}
};
\end{tikzpicture}%

\subsection{Result for RQ2: Fault Detection Capability}
\label{sec:rq2}

\begin{figure*}[t]
\centering
\begin{subfigure}[b]{0.42\textwidth}
	\includegraphics[width=\columnwidth]{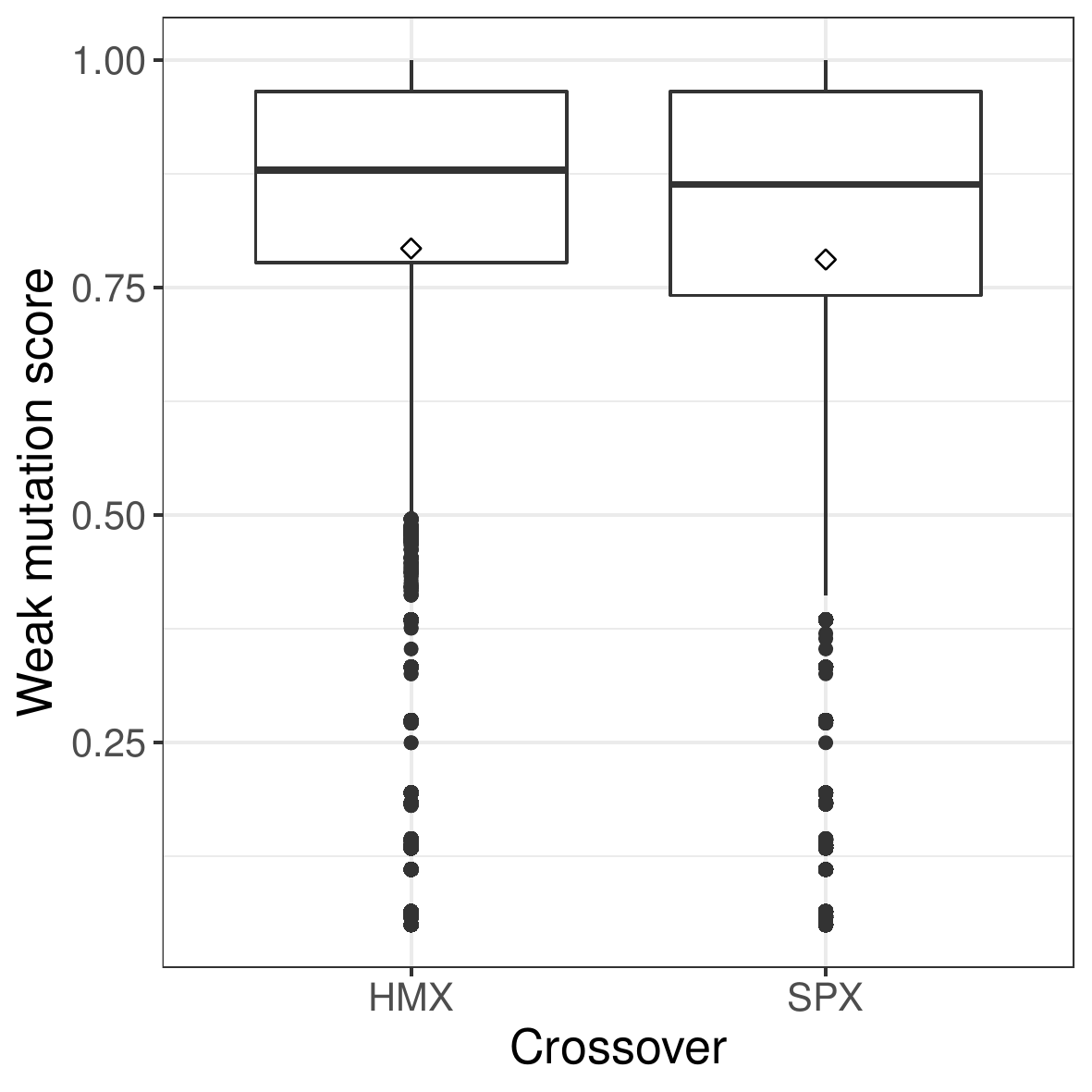}
	\caption{Weak Mutation Score}
	\label{fig:mutation-weak}
\end{subfigure}
\begin{subfigure}[b]{0.42\textwidth}
	\includegraphics[width=\columnwidth]{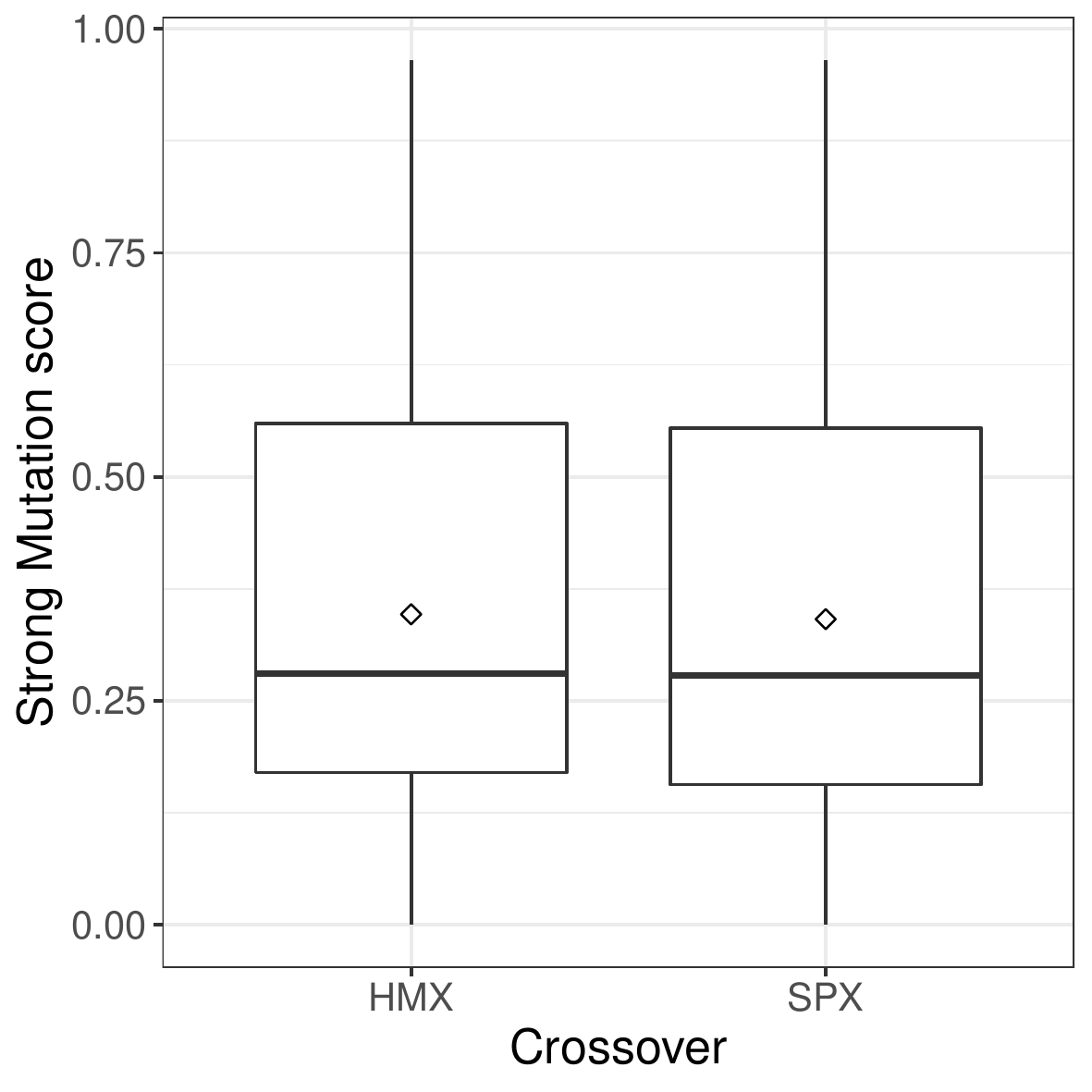}
	\caption{Strong Mutation Score}
	\label{fig:mutation-strong}
\end{subfigure}
\caption{Boxplot of structural coverage comparing \newcrossover{} to the baseline SPX.}
\label{fig:mutation}
\end{figure*}

\cref{fig:mutation} shows the fault detection capability of \newcrossover{} compared to SPX measured through the mutation score.
\cref{fig:mutation-weak} shows the weak mutation score and \cref{fig:mutation-strong} shows the strong mutation score.
The boxplots show the median, quartiles, variability in the results, and the outliers for all classes in the benchmark together.
The diamond point indicates the mean of the results.
From \cref{fig:mutation-weak}, we can see that, on average, \newcrossover{} improves the weak mutation score by +\SI{1.2}{\percent} compared to SPX.
However, from \cref{fig:mutation-strong} we can see that overall, the strong mutation scores only show marginal improvements (+\SI{0.5}{\percent}).

\begin{table}[t]
	\centering
	\caption{Statistical results of \newcrossover{} vs. SPX for fault-detection capability.}
	\scriptsize
	\begin{tabular}{lc cccc c cccc  cc}
		\toprule
		\multirow{2}{*}{Metric} && \multicolumn{4}{c}{\#Win} && \multicolumn{4}{c}{\#Lose} && \multirow{ 2}{*}{\#No diff.} \\
		\cmidrule{2-6} \cmidrule{8-11}&& Negl. & Small & Medium & Large && Negl. & Small & Medium & Large \\
		\midrule
		Weak mutation && 3 & 3 & 3 & 21 && 0 & 1 & 0 & 0 && 72\\
		Strong mutation && 0 & 8 & 0 & 15 && 0 & 3 & 0 & 0 && 77\\
		\bottomrule
	\end{tabular}
	\label{tab:mutation-statistics}
\end{table}

\cref{tab:mutation-statistics} shows the statistical comparison between \newcrossover{} and SPX, based on a $\pvalue{} \leq 0.05$.
Similarly to \cref{tab:structural-statistics}, \textit{\#Win} indicates the number of times that \newcrossover{} has a statistically significant improvement over SPX, \textit{\#Equal} indicates the number of times that there is no statistical difference in the results of the two operators, and \textit{\#Lose} indicates the number of times that \newcrossover{} has statistically worse results than SPX.
The \textit{\#Win} and \textit{\#Lose} columns additionally also indicate the magnitude of the difference through the $\atwelve{}$ effect size.
From \cref{tab:mutation-statistics}, we can see that \newcrossover{} has a statistically significant non-negligible improvement in 27 and 23 cases for weak and strong mutation, respectively.
For weak mutation, \newcrossover{} improves with a large magnitude for 21 classes, medium for 3 classes, and small for 3 classes.
For strong mutation, \newcrossover{} improves with a large magnitude for 15 classes and a small magnitude for 8 classes.
\newcrossover{} performes worse in one case (\texttt{Fraction} from the \texttt{Lang} project) for weak mutation and three cases (\texttt{AdaptiveStepsizeFieldIntegrator} and \texttt{MultistepIntegrator} from the \texttt{Math} project, and \texttt{SphericalCoordinates} from the \texttt{Geometry} project) for strong mutation, all with a small effect size.

We observe a maximum increase in weak mutation score of +\SI{14.0}{\percent} for the \texttt{hungarianStemmer} class (\texttt{Stemmer}) and +\SI{12.2}{\percent} for the \texttt{ExtendedMessageFormat} class (\texttt{Text}) on strong mutation score.
Among the classes that improve on weak and strong mutation score, 27 and 20, respectively, also improve \wrt{} branch coverage.
Interestingly, four classes among both mutation scores improve \wrt{} mutation score without improving the structural coverage.

\vspace*{2mm}
\hspace*{-5mm}
\begin{tikzpicture}
\node [mybox] (box){%
\centering
\begin{minipage}{0.94\textwidth}
In summary, \newcrossover{} achieves significantly higher (\textasciitilde\SI{25}{\percent} of the cases) or equal fault detection capability compared to SPX and is outperformed in one and three classes for weak and strong mutation, respectively.
\end{minipage}
};
\end{tikzpicture}%

\section{Threats to Validity}
\label{sec:validity}

This section discusses the potential threats to the validity of our study.

\textbf{Construct validity}: Threats to \textit{construct validity} stem from how well the chosen evaluation metrics measure the intended purpose of the study.
Our study relies on well-established evaluation metrics in software testing to compare the proposed hybrid multi-level crossover with the current state-of-the-art, namely  structural coverage (\ie{} branch and line) and fault detection capability (\ie{} weak and strong mutation).
As the stopping condition of the search process, we used a time-based budget rather than a budget based on the number of test evaluations or generations.
A time-based budget provides a fairer measure since the two crossover operators have a different overhead and execution time and might otherwise provide an unfair advantage to our operator.

\textbf{Internal validity}: Threats to \textit{internal validity} stem from the influence of other factors onto our results.
The only difference between the two approaches in our study is the crossover operator.
Therefore, any improvement or diminishment in the results must be attributed to the difference in the two crossover operators.

\textbf{External validity}: Threats to \textit{external validity} stem from the generalizability of our study.
We selected \num{116}~classes from popular open-source projects based on their cyclomatic complexity and type of input parameters to create a representative benchmark.
These classes have previously been used in the related literature on test case generation~\cite{panichella2015reformulating, panichella2017automated}.

\textbf{Conclusion validity}: Threats to \textit{conclusion validity} stem from the deduction of the conclusion from the results.
To minimize the risk of the randomized nature of EAs, we performed \num{100}~iterations of the experiment in our study with different random seeds.
We have followed the recommended guidelines for running empirical experiments with randomized algorithms using sound statistical analysis as recommend in the literature~\cite{arcuri2014hitchhiker}.
We used the unpaired Wilcoxon rank-sum test and the Vargha-Delaney $\atwelve{}$ effect size to determine the significance and magnitude of our results.

\section{Conclusions and Future Work}
\label{sec:conclusion}

In this paper, we have proposed a novel crossover operator, called \newcrossover{}, that combines different crossover operators on both a test case-level and a data-level for generating unit-level test cases.
By implementing such a hybrid multi-level crossover operator,  we can create genetic variation in not only the test statements but also the test data.
We implemented \newcrossover{} in \evosuite{}, a state-of-the-art Java unit test case generation tool.
Our approach was evaluated on a benchmark of \num{116}~classes from two popular open-source projects.
The results show that \newcrossover{} significantly improves the structural coverage and fault detection capability of the generated test cases compared to the standard crossover operator used in \evosuite{} (\ie{} single-point).
Based on these promising results, there are multiple potential directions for future work to explore.
In this paper, we detailed the crossover operator for two types of primitive test data inputs (\ie{} numbers and strings).
In future work, we are planning to extend this with additional operators for arrays, lists, and maps.
Additionally, we want to experiment with alternative crossover operators for numbers (\eg{} parent-centric crossover, arithmetic crossover) and strings (\eg{} multi-point crossover).

\section*{Acknowledgements} 
We gratefully acknowledges the Horizon 2020 (EU Commission) support for the project \textit{COSMOS}, Project No. 957254-COSMOS.

\bibliographystyle{splncs04}
\bibliography{bibliography}
\end{document}